\documentstyle[12pt,a41,psfig,here]{article}

\newcommand{\be}{\begin{equation}}
\newcommand{\ee}{\end{equation}}
\newcommand{\bea}{\begin{eqnarray}}
\newcommand{\eea}{\end{eqnarray}}
\newcommand{\ba}{\begin{array}}
\newcommand{\ea}{\end{array}}

\newcommand{\slashl}[1]{\not{\!\!#1}}

\begin{document}

\begin{center}
{\LARGE\bf Renormalization of the Twist-3
Flavor Singlet Operators in a Covariant Gauge}

\vspace{1cm}
{J. K{\scriptsize ODAIRA}$^a$,  T. N{\scriptsize ASUNO}$^a$,  
H. T{\scriptsize OCHIMURA}$^a$,  
K. T{\scriptsize ANAKA}$^b$,  
and Y. Y{\scriptsize ASUI}$^c$}

\vspace*{1cm}
{\it $^a$Department of Physics, Hiroshima University, 
Higashi-Hiroshima 739, Japan}\\

\vspace*{3mm}
{\it $^b$Department of Physics, Juntendo University, Inba-gun, 
Chiba 270-16, Japan}\\

\vspace*{3mm}
{\it $^c$RIKEN BNL Research Center, Brookhaven National Laboratory, 
Upton, NY, 11973, USA}\\

\vspace*{2cm}

\end{center}

\begin{abstract}
We investigate the nucleon's transverse spin-dependent structure
function $g_{2}(x, Q^{2})$
in the framework of the operator product expansion and the 
renormalization group.
We construct the complete set of the twist-3 operators 
for the flavor singlet channel, and give the relations among them.
We develop an efficient, covariant approach
to derive the anomalous dimension matrix 
of the twist-3 singlet operators
by computing the off-shell
Green's functions.  
As an application, we investigate
the $Q^{2}$-evolution of $g_{2}(x, Q^{2})$
for the lowest moment
case, and discuss its experimental implication.
\end{abstract}

\section{Introduction}

\vspace{1mm}
\noindent
The nucleon spin structure observed in deep inelastic scattering (DIS)
%which is characterized by large momentum transfer squared $Q^{2}$
%and the Bjorken variable $x$,
is described by the two structure functions $g_{1}(x, Q^{2})$
and $g_{2}(x, Q^{2})$. 
$g_{1}$ and $g_{2}$ are present for the longitudinally and 
transversely polarized targets, respectively.
In the framework of the operator product expansion, 
the leading contribution for $g_{1}$ comes only from the 
twist-2 operators, while for $g_{2}$ the twist-3 operators
also contribute in the leading order of $1/Q^{2}$ in
addition to the twist-2 operators\cite{HM,JA}.
%\cite{HM,AR,KMSU,JA}.
$g_{2}(x, Q^{2})$ 
is unique, and 
plays a distinguished role in spin physics\cite{HM,JA,SV}.
Firstly, it is a ``measurable'' higher-twist structure
function:
%\cite{HM,JA}: 
In general, it is difficult to extract the higher-twist
structure functions by experiments because they usually constitute
small corrections to the leading twist-2 term. However,
$g_{2}(x, Q^{2})$ is somewhat immune to this difficulty, and
contributes as a leading term to the asymmetry in the DIS
using the transversely polarized target.
Secondly, $g_{2}(x, Q^{2})$ is related to an interesting sum rule, 
referred to as
the Burkhardt-Cottingham sum rule:
%\cite{BK}: 
$\int_{0}^{1} dx g_{2} (x, Q^{2})
= 0$.
Thirdly,  
$g_{2}(x, Q^{2})$ contains information inaccessible by the more familiar
spin structure function $g_{1}(x, Q^{2})$: 
$g_{2}(x, Q^{2})$ is related to the nucleon's transverse polarization 
and to the twist-3 operators describing the quark-gluon 
correlation in the nucleon.
%\cite{SV,JA}.

Recently, the first data of $g_{2}(x, Q^{2})$ have been 
reported\cite{S143}.
% by measuring spin
%asymmetry in the scattering of the longitudinally polarized leptons
%off the transversely polarized nucleon.
Extensive study is expected to be performed at DESY, CERN, 
and SLAC. These future measurements of $g_{2}(x, Q^{2})$
will give the first opportunity
to obtain new information about the dynamics of QCD
and the nucleon structure beyond those from the twist-2
structure functions. 
Theoretically, the determination of the $Q^{2}$-dependence of 
$g_{2}(x, Q^{2})$ is very important to extract these physical
information from experimental data.
Furthermore, the comparison of the $Q^{2}$-evolution itself
between theory and experiment
will provide a deeper test of QCD
beyond the conventional twist-2 level.

The $Q^{2}$-evolution of $g_{2}(x, Q^{2})$ is governed by the
anomalous dimensions, which are determined by the renormalization of
the relevant twist-3 operators as well as the twist-2 ones.
A characteristic feature of the higher-twist operators is the
occurrence of the complicated operator mixing under renormalization:
Many gauge-invariant operators, the number of which increases with spin
(moment of the structure function), 
mix with each other.\footnote{Recently, it has been proved
that the twist-3 structure functions obey simple DGLAP
evolution equation for $N_{c} \rightarrow \infty$,
and that this simplification
is universal for all twist-3 nonsinglet
structure functions\cite{ABH}.}
%See also \cite{BEL}.}
Furthermore, the operators which are proportional to the equation of
motion (``EOM operators''), as well as the ones which are not gauge
invariant (``alien operators''), 
also participate through the renormalization mixing \cite{POLI,JCOL1}.

There have been a lot of works on the $Q^{2}$-evolution of
$g_{2}(x,Q^{2})$.
Most of them discussed the flavor nonsinglet case \cite{RJC,KOD1,KOD2}.
Only a few works treated the singlet case\cite{BKL}:
Bukhvostov, Kuraev and Lipatov 
%\cite{BKL} 
derived evolution equations
for the twist-3 quasi-partonic operators.
Recently, M\"uller 
%\cite{DMU} 
computed evolution kernel based on the
nonlocal light-ray operator technique, and obtained the identical
results.
%to \cite{BKL}.
However, both of these two works employ a similar framework based on
the renormalization of the nonlocal operators in the (light-like) axial 
gauge.
Balitsky and Braun
%\cite{BB} 
also treated the nonlocal operators
although they employed the background field method. 
On the other hand, covariant
approach based on the local composite operators
is missing.
Furthermore, some subtle infrared problem occurring in the
renormalization of the generic flavor singlet operators has been emphasized
in \cite{JCOL2}.
Therefore, the computation of the anomalous dimensions for the
flavor singlet part
in a covariant and fully consistent scheme is desirable and should provide
useful framework.

In this work, we develop a covariant framework to
investigate the flavor singlet part of $g_{2}(x, Q^{2})$
based on the operator product expansion (OPE) 
and the renormalization group (RG),
by extending our recent work on the flavor nonsinglet 
part\cite{KOD1,KOD2}.
%In this work we analyze the twist-3 flavor singlet operators for
%$g_{2}(x, Q^{2})$, which must be important from both theoretical and
%phenomenological viewpoints.
Sect.2 is devoted to a detailed OPE analysis for the flavor singlet
part of $g_{2}(x, Q^{2})$.
We list up all relevant twist-3 flavor singlet operators
appearing in QCD.
We derive the relations satisfied by these operators.
In particular, we obtain a new operator identity,
which relates the gluon bilinear operator
with the trilinear ones.
Based on these developments, we give a basis of the independent
operators for the renormalization.
In sect.3, we discuss a general framework to perform
the renormalization of the twist-3 flavor singlet operators
in a covariant gauge.
We compute the off-shell Green's functions inserting
the relevant local composite operators.
Infrared cut-off is provided by the external off-shell momenta.
In this case, the EOM operators as well as the alien operators
should be included as independent operators.
We introduce convenient projection technique\cite{TK,KOD2}
to avoid the complexity stemmed from the renormalization mixing
of many gauge-noninvariant operators.
%Recently, a convenient technique in this scheme has been developed and 
%efficiently applied to the flavor non-singlet case of $g_{2}(x, Q^{2})$
%\cite{KOD1,KOD2}.
%Here we extend the method to the singlet case.
Sect.4 contains an application of our framework
to investigate the $Q^{2}$-evolution of the lowest ($n=3$) moment.
We discuss experimental as well as theoretical implication of our results.
%Some results for the case of the general moment $n$ are also
%mentioned.
The final sect.5 concludes with a brief summary.

\section{OPE analysis of twist-3 flavor singlet operators for 
$g_{2}(x, Q^{2})$}

\vspace{1mm}
\noindent
%In this section we present a detailed OPE analysis 
%to give the complete set of the twist-3 flavor singlet operators
%for $g_{2}(x, Q^{2})$.
%We follow the convention of \cite{KOD1,KOD2}.
%
Let us start from a quick look at the factorization theorem
in QCD\cite{POLI,CSS}. As is well known,
the cross section for the DIS
%,  $l + H \rightarrow  l' + X$,
is given by the ``cut diagram'' 
corresponding to the discontinuity of the
forward virtual Compton amplitude
between the virtual photon with the momentum
$q_{\mu}$ ($q^{2} = - Q^{2}$) and
the nucleon with the momentum $P_{\mu}$
($P^{2} = M^{2}$ with $M$ the nucleon mass).
In general kinematics, 
%the blob of fig.\ref{fig1}
this virtual Compton amplitude
contains all the complicated interactions
between the virtual photon and the nucleon, possibly
including the ``soft interactions'' where the soft
momenta are exchanged.
However, drastic simplification occurs if one
goes to the Bjorken limit
$Q^{2} \rightarrow \infty$ with $x = Q^{2}/2P\cdot q$ fixed:
The amplitude is dominated by the contribution which
is factorized into the hard (short distance) and the soft 
(long distance) parts, and the other
complicated contributions are suppressed by the powers of $1/Q$.
The factorized amplitude corresponds to the process
where a parton carrying the momentum $k=\xi P$
($0 \leq \xi \leq 1$) comes out from the soft part,
followed by the hard hitting by the virtual photon,
and then goes back to the soft part. 
For the case of the polarized target,
the structure functions $g_{1}(x, Q^{2})$
and $g_{2}(x, Q^{2})$ appear in the cross section,
and they are given by, 
corresponding to the
factorized amplitude,
\begin{eqnarray}
g_{1}(x, Q^{2}) &=& \sum_{i}
\int_{x}^{1} \frac{d\xi}{\xi} 
\phi^{i}_{1}(\xi, \mu^{2}) H_{1i}\left(\frac{x}{\xi},
\frac{Q}{\mu}, \alpha_{s}(\mu)\right), 
\label{eq:g1} \\
g_{2}(x, Q^{2}) &=& \sum_{i}
\int_{x}^{1} \frac{d\xi}{\xi} 
\phi^{i}_{2}(\xi, \mu^{2}) H_{2i}\left(\frac{x}{\xi},
\frac{Q}{\mu}, \alpha_{s}(\mu)\right).
\label{eq:g2}
\end{eqnarray}
where $\phi^{i}_{1}, \phi^{i}_{2}$ are
the parton distribution
functions corresponding to the soft part.
As is well known, 
the twist-2 distribution $\phi^{i}_{1}(\xi, \mu^{2})$ is 
interpreted as the probability density to
find a parton of type $i$ ($={\rm gluon},
u, \overline{u}, d, \overline{d}, \cdots$) in the nucleon,
carrying a fraction $\xi$ of the nucleon's momentum.
The summation of (\ref{eq:g1}), (\ref{eq:g2}) is over all the
possible types of parton, $i$.
$H_{1i}, H_{2i}$ denote the hard parts;
they correspond to the hard scattering coefficients
for the scattering
between the virtual photon and a parton $i$.
%corresponding to the hard part.
They are calculable systematically by perturbation theory, and
the dependence on the strong
coupling constant $\alpha_{s}=g^{2}/4\pi$ is explicitly shown.
%The equation (\ref{eq:f2}) can be proved by analyzing
%directly the Feynman graphs
%for the forward Compton amplitude in the Bjorken limit\cite{CSS}.
%In the present case of the DIS, the equivalent result
%can be obtained by the operator product
%expansion\cite{Muta,JC2}.
In (\ref{eq:g1}), (\ref{eq:g2}),
the soft and the hard parts 
are divided at the renormalization scale $\mu$.

The soft part (parton distribution function)
is a process-independent quantity
which appears universally in various hard processes.
Following \cite{JJ}
%\cite{CS,JJ},
this is written
as Fourier transform of nucleon matrix elements
of nonlocal light-cone operators,
reflecting the light-cone dominance in hard processes.
For our case, the relevant quantity is
\begin{equation}
\Phi_{\mu}^{i}(\xi, \mu^{2}) =
%\int
%\frac{d^{4}k}{(2\pi)^{4}} \delta ( k^{+}
%- xP^{+} )
P^{+} \int \frac{dz^{-}}{2\pi}
e^{i\xi P\cdot z} \langle P S|
\overline\psi_{i}(0) \gamma_{\mu}\gamma_{5}[0, z] 
\psi_{i}(z) |P S \rangle.
\label{eq:gam}
\end{equation}
Here $z$ is a light-like vector $z^{2} = 0$,
$|P S \rangle$ is the nucleon state
with momentum $P$ and spin $S$
($P\cdot S = 0, S^{2}= - M^{2}$),
and $\psi_{i}$ is the quark field.
Our Lorentz frame is chosen as $P \cdot z = P^{+}z^{-}$.
The bilocal operator is renormalized at the scale $\mu$.
$\left[0, z\right] = {\rm P} 
\exp \left[ ig \int_{1}^{0}d\alpha z^{\mu}
A_{\mu}(\alpha z) \right]$ 
is the path-ordered gauge factor, and
therefore (\ref{eq:gam}) is gauge-invariant.
We considered the case where the quark of flavor $i$
comes out of the nucleon, followed by the
coupling with the vertex $\gamma_{\mu} \gamma_{5}$, 
and then goes back
to the nucleon. The quark has $\xi P^{+}$
as the plus component of the momentum.
%Several comments are in order here:
%(i) Here and in the following,
%$z^{\mu}$ denotes the light-like vector as
%$z_{\mu}z^{\mu} = 0$, $z^{+} = \bold{z}_{\perp} =0$.
%(The light-cone coordinates
%$a^{\mu} = (a^{+}, a^{-}, \bold{a}_{\perp})$
%are related to the usual coordinates
%$a^{\mu} = (a^{0}, a^{1}, a^{2}, a^{3})$ by
%$a^{\pm} =
%(a^{0} \pm a^{3})/\sqrt{2}$,
%$\bold{a}_{\perp} = (a^{1}, a^{2})$\cite{KS}.)
%Thus, (\ref{eq:gam}) is
%the light-cone Fourier transform of the matrix elements of
%the nonlocal operator.
%This reflects the light-cone dominance of the hard processes.

In principle, one can insert various gamma matrices
in between the quark fields.
In fact, one can generate all quark distribution functions
up to twist-4 by inserting 
all the possible gamma matrices\cite{JJ}.
For example, if $\gamma_{\mu}$ were inserted
in place of $\gamma_{\mu}\gamma_{5}$, we would
obtain the spin-independent distribution functions
of twist-2 and twist-4,
corresponding to the structure functions $F_{1}(x, Q^{2})$,
$F_{2}(x, Q^{2})$.
In our case of (\ref{eq:gam}), additional $\gamma_{5}$
distinguishes the helicity of the quark. Therefore,
(\ref{eq:gam}) gives the spin-dependent distribution functions
$\phi_{1}^{i}$, $\phi_{2}^{i}$ relevant to the structure 
functions $g_{1}(x, Q^{2})$ and $g_{2}(x, Q^{2})$.
Explicitly, we have\cite{JJ}
\begin{equation}
\Phi_{\mu}^{i}(\xi, \mu^{2}) = 2 \left[\phi_{1}^{i}(\xi, \mu^{2})
P_{\mu}\frac{(S \cdot z)}{(P\cdot z)}
+ \left( \phi_{1}^{i}(\xi, \mu^{2}) + \phi_{2}^{i}(\xi, \mu^{2})
\right) S_{\perp \mu}\right],
\label{eq:sep}
\end{equation}
up to the twist-4 corrections. Here $S_{\perp \mu}
= S_{\mu} - P_{\mu} (S \cdot z)/(P\cdot z)
+ M^{2} z_{\mu} (S \cdot z)/(P \cdot z)^{2}$ is the projection
of the spin vector
into transverse direction. 
By comparing both sides of (\ref{eq:sep}), one can obtain
explicit operator definitions of the relevant distribution 
functions.

For further development, it is convenient to employ the moment
given by $\int d\xi \xi^{n-1}\Phi_{\mu}^{i}(\xi, \mu^{2})$,
which is related to the moments of the structure functions
$M_{n}[g_{1}(Q^{2})] \equiv \int_{0}^{1} dx x^{n-1} g_{1}(x, Q^{2})$,
$M_{n}[g_{2}(Q^{2})] \equiv \int_{0}^{1} dx x^{n-1} g_{2}(x, Q^{2})$
(see (\ref{eq:g1}), (\ref{eq:g2})):
As seen from (\ref{eq:gam}), going over to the moment space
corresponds to expanding the nonlocal operator
at small quark-antiquark separation.
This gives the OPE.
We obtain (we consider the case of odd $n$ which is relevant
for the DIS)
\begin{eqnarray}
\int^{1}_{-1} d\xi \xi^{n-1} \Phi_{\mu}^{i}(\xi, \mu^{2})
&=& \int_{0}^{1} d\xi \xi^{n-1} \left(
\Phi_{\mu}^{i} (\xi, \mu^{2}) + \Phi_{\mu}^{\overline{i}} 
(\xi, \mu^{2})\right)
\nonumber \\
&=& \Delta^{\mu_{1}} \cdots \Delta^{\mu_{n-1}}
\langle PS | \overline{\psi}_{i}(0) \gamma_{\mu}\gamma_{5}
iD_{\mu_{1}} \cdots iD_{\mu_{n-1}} \psi_{i}(0) |PS \rangle,
\label{eq:expa}
\end{eqnarray}
where $\Delta^{\mu} = z^{\mu}/(P \cdot z)$, and 
$D_{\mu} = \partial_{\mu} - igA_{\mu}$ is
the covariant derivative.
The local composite operators in the r.h.s.
are symmetric and traceless for the indices $\mu_{1}, \cdots, \mu_{n-1}$,
but have one free index $\mu$. If one recalls that the twist
of the local composite operator is defined as
``dimension minus spin'', one immediately concludes the following:
The symmetrization of the index $\mu$ with the other indices
and the subtraction of the trace terms corresponding to
$g_{\mu \mu_{k}}$ give the twist-2 operators; these twist-2
operators correspond to $g_{1}(x, Q^{2})$.
On the other hand, the antisymmetrization
of $\mu$ with the other indices introduces
one unit of twist, giving the twist-3 operators; the matrix
element of these operators, having antisymmetric pair
of indices, should involve $S_{\perp \mu}$, and therefore 
contributes to $g_{2}(x, Q^{2})$ (see (\ref{eq:sep})).

From (\ref{eq:sep}) and (\ref{eq:expa}), one sees that the twist-2 
operators also contribute to $g_{2}(x, Q^{2})$. As is well known,
these contributions can be conveniently extracted by\cite{JA}
%\cite{WW}
\begin{equation}
g_{2}(x, Q^{2}) = - g_{1}(x, Q^{2}) + \int^{1}_{x} dy 
\frac{g_{1}(y, Q^{2})}{y} + \widetilde{g}_{2}(x, Q^{2}),
\label{eq:ww}
\end{equation}
where we denote the twist-3 part by
$\widetilde{g}_{2}(x, Q^{2})$. 
%The twist-2 contributions
%to $g_{2}(x, Q^{2})$ are completely determined by
%information on $g_{1}(x, Q^{2})$.
Now $M_{n}[\widetilde{g}_{2}(Q^{2})]$ is given by matrix element
of 
\begin{equation}
  R_{n,F}^{\sigma\mu_{1}\cdots \mu_{n-1}} =
         i^{n+1}\frac{2(n-1)}{n} {\cal S}{\cal A} {\cal S}
\left[ \overline{\psi}
       \gamma^{\sigma}\gamma_5
D^{\mu_1} \cdots D^{\mu_{n-1}}\psi \right] - {\rm traces},
\label{quark}
%           \nonumber\\
%   & & \qquad - \sum_{l=1}^{n-1} \overline{\psi} \gamma_5
%       \gamma^{\mu_l }D^{\{\sigma} D^{\mu_1} \cdots D^{\mu_{l-1}}
%            D^{\mu_{l+1}} \cdots D^{\mu_{n-1}\}}
%             \psi \Bigr] - {\rm traces} \ , \label{quark}\\
\end{equation}
where $\cal{S}\mit$ symmetrizes $\mu_{1},\cdots,\mu_{n-1}$, and
$\cal{A}\mit$ antisymmetrizes $\sigma$ and $\mu_{1}$
as ${\cal A}f^{\sigma \mu_{1}} = \frac{1}{2} (f^{\sigma \mu_{1}}
- f^{\mu_{1} \sigma})$.
``$-$ traces'' stands for the subtraction of
the trace terms to make the operators traceless. 
$n= 3, 5, 7, \cdots$, because only the charge-conjugation even
operators contribute to the DIS.
Here and below we suppress the flavor index of the quark field:
For the flavor nonsinglet part, the flavor 
matrices $\lambda_{i}$ should be inserted between the quark fields.
On the other hand, for the present case of the singlet part,
the flavor indices have to be summed over.

In addition to (\ref{quark}),
one can construct 
other types of gauge-invariant
operators of twist-3\cite{SV,RJC}:
%\cite{SV,RJC,KOD1,KOD2}:
%
\begin{eqnarray}
  R_{n,l}^{\sigma \mu_{1} \cdots \mu_{n-1}} &=& 
     \frac{1}{2n}\left(V_{l}-V_{n-1-l} + U_{l} + U_{n-1-l} \right)
     \;\;\;\;\;\;\;(l = 1, \cdots, n-2) \ , \label{eq:rl}\\
  R_{n,m}^{\sigma \mu_{1} \cdots \mu_{n-1}}&=&i^{n}{\cal S}m
          \overline{\psi}\gamma^{\sigma}\gamma_{5}
           D^{\mu_{1}}\cdots D^{\mu_{n-2}}\gamma^{\mu_{n-1}}\psi
             - {\rm traces} \ , \label{eq:rm}\\
  R_{n,E}^{\sigma \mu_{1} \cdots \mu_{n-1}} &=& i^{n}\frac{n-1}{2n}
   {\cal S} \left[ \overline{\psi}(i \slashl{D} - m ) 
  \gamma^{\sigma} \gamma_{5} D^{\mu_{1}} \cdots D^{\mu_{n-2}}
        \gamma^{\mu_{n-1}} \psi \right.\nonumber \\
     & & \qquad + \,\left.\overline{\psi} \gamma^{\sigma}\gamma_5 
    D^{\mu_{1}} \cdots D^{\mu_{n-2}}\gamma^{\mu_{n-1}}(i\slashl{D}
      - m ) \psi \right] - {\rm traces} \ , \label{eq211}
\end{eqnarray}
where
\begin{eqnarray}
   V_{l}&=&i^{n} g {\cal S} \overline{\psi}\gamma_{5} 
       D^{\mu_1} \cdots G^{\sigma \mu_l} \cdots
       D^{\mu_{n-2}}\gamma^{\mu_{n-1}} \psi - {\rm traces},
\label{eq:v}\\
   U_{l}&=&i^{n-3} g {\cal S}\overline{\psi}
      D^{\mu_1} \cdots \widetilde{G}^{\sigma \mu_l} \cdots
      D^{\mu_{n-2}}\gamma^{\mu_{n-1}} \psi - {\rm traces}.
\label{eq:u}
\end{eqnarray}
Here
$m$ represents the quark mass (matrix).
The combination in the r.h.s. of (\ref{eq:rl})
constitutes an independent set of 
charge-conjugation even operators. 
The operators in (\ref{eq:rl}) contain
the gluon field strength $G_{\mu\nu}$ or the dual tensor
$\widetilde{G}_{\mu \nu}={1\over
2}\varepsilon_{\mu\nu\alpha\beta}
G^{\alpha\beta}$ explicitly; this implies
that they represent the effect of quark-gluon correlations.
$R_{n,E}$ is the EOM operator;
it vanishes by the use of the QCD equations of motion,
although it is in general a nonzero operator
due to quantum effects.
$R_{n,m}$ is due to the quark mass effect.

The above operators are not all independent. By using
$D_{\mu} = \{ \gamma_{\mu}, \not{\!\!\! D} \}/2$ and
$[D_{\mu}, D_{\nu}] = -igG_{\mu \nu}$,
it is straightforward to derive
the following equation\cite{SV,JA}:
\begin{equation}
 R_{n,F}^{\sigma\mu_{1}\cdots \mu_{n-1}} = 
        \frac{n-1}{n} R_{n,m}^{\sigma\mu_{1}\cdots \mu_{n-1}}
             + \sum_{l=1}^{n-2} (n-1-l)
                 R_{n,l}^{\sigma\mu_{1}\cdots \mu_{n-1}} +
             R_{n,E}^{\sigma\mu_{1}\cdots \mu_{n-1}} \ . 
\label{operel}
\end{equation}
Therefore we can exclude one operator
among (\ref{quark})-(\ref{eq211}) to form an independent basis.
A convenient choice of the independent operators will be 
(\ref{eq:rl}), (\ref{eq:rm}) and (\ref{eq211}).

For the flavor nonsinglet part of $g_{2}(x, Q^{2})$,
the operators (\ref{eq:rl})-(\ref{eq211}) form 
the complete set of the twist-3 gauge-invariant operators.
For the singlet part, however, this is not the whole story:
The QCD radiative corrections could replace the quark fields
involved in (\ref{quark}), (\ref{eq:rl}), (\ref{eq:v})
and (\ref{eq:u}) by the gluon fields, and generate the operators
which are bilinear or trilinear in the gluon fields.
In addition, because the QCD lagrangian
in a covariant gauge involves the ghost field
as a dynamical variable,
the alien operators involving these unphysical degrees of freedom
could also participate.
These considerations lead to
the following ``new'' operators:
%\vspace*{-1pc}
\bea
T_{n,G}^{ \sigma \mu_{1} \cdots \mu_{n-1} }
& = &
i^{ n-1 }
\mbox{$ \cal{S} \mit $} \mbox{ $ \cal{A} \mit $ } 
\mbox{$ \cal{S} \mit $} \left[ \widetilde{G}^{ \nu \mu_{1} } 
D^{ \mu _{2} } \cdots D^{ \mu _{n-1} } G_{ \nu }^{ ~ \sigma }
\right]
- \mbox{traces},
\label{eqn:1}
\\
T_{n,l}^{ \sigma \mu_{1} \cdots \mu_{n-1} }
& = &
i^{n-2} g 
\mbox{ $ \cal{ S } \mit $ } \left[ 
{G}^{\nu \mu _{1}}
D^{ \mu _{2} } \cdots \widetilde{G}^{ \sigma \mu _{l}}
\cdots D^{ \mu _{n-2} } G_{ \nu }^{ ~ \mu _{n-1} }
\right]
- \mbox{traces}~~~(l=2,\ldots,\frac{n-1}{2}),~
\label{eqn:2}
\\
\widetilde{T}_{n,l}^{ \sigma \mu_{1} \cdots \mu_{n-1} }
& = &
i^{n-2} g 
\mbox{ $ \cal{ S } \mit $ } \left[ 
\widetilde{G}^{\nu \mu _{1}}
D^{ \mu _{2} } \cdots {G}^{ \sigma \mu _{l}}
\cdots D^{ \mu _{n-2} } G_{ \nu }^{ ~ \mu _{n-1} }
\right]
- \mbox{traces}~~~(l=2,\ldots,\frac{n-1}{2}),~
\label{eqn:2t}
\\
T_{n,B}^{ \sigma \mu_{1} \cdots \mu_{n-1} }
& = &
i^{n-1} {\cal S}\{ \widetilde{G}^{ \sigma \mu_{1} } D^{ \mu _{2} }
\cdots D^{ \mu _{n-2} }
\}^{a}
\{ 
- \frac{1}{ \alpha } \partial ^{ \mu_{n-1} }(  \partial ^{ \nu }
A^{a}_{ \nu } )
+
g f^{abc}(  \partial ^{ \mu_{n-1}} \overline{\chi}^{ b } ) \chi ^{c}
\} - \mbox{traces},
\label{eqn:3}
\\ 
T_{n,E}^{ \sigma \mu_{1} \cdots \mu_{n-1} }
& = &
i^{n-1} {\cal S}\{ \widetilde{G}^{ \sigma \mu_{1} } D^{ \mu _{2} }
\cdots D^{ \mu _{n-2} }
\}^{a}
\{ 
\left ( D^{ \nu }G_{ \nu }^{~ \mu_{n-1} } \right )^{a}
\nonumber \\
& \mbox{} &
+
g \overline{ \psi } t^{a} \gamma^{ \mu_{n-1} } \psi
+
\frac{1}{ \alpha } \partial ^{ \mu_{n-1} }(  \partial ^{ \nu }
A^{a}_{ \nu })
-
g f^{abc}(  \partial ^{ \mu_{n-1}} \overline{ \chi } ^{ b } ) \chi ^{c}
\}  -\mbox{traces}.
\label{eqn:4}
\eea
Here the gluon field $A_{ \mu }$ and the covariant derivative 
$D^{\mu}$ are in the adjoint representation.
$\chi$ and $\overline{\chi}$ are the ghost fields, and
$\alpha$ is the gauge parameter.
%$\alpha=1$ in the Feynman gauge.
$t^{a}$ is the color matrix as 
$[t^{a}, t^{b}]=i f^{abc} t^{c}$, 
${\rm Tr}(t^at^b) = {1 \over 2}\delta^{ab}$.
$T_{n,l}$ and $\widetilde{T}_{n,l}$ 
are trilinear in the gluon field strength and its dual
tensor,
%$G_{ \mu \nu }$ and 
%the dual tensor $\tilde{G}_{\mu \nu}=(1/2) \varepsilon_{ \mu \nu
%\alpha \beta}G^{ \alpha \beta }$, 
and thus represent the effect of three
gluon correlations.
In contrast to the quark-gluon operators
(\ref{eq:rl}), (\ref{eq:v}) and (\ref{eq:u}),
$T_{n,l}$ ($\widetilde{T}_{n,l}$) is charge-conjugation even
by itself.
$T_{n,E}$ is the EOM operator,
and vanishes by the naive use of the equations of motion
for the gluon. 
$T_{n,B}$ is the BRST-exact
operator \cite{JCOL1} which is the BRST variation of the operator:
$i^{n-1} {\cal S}\{ \widetilde{G}^{ \sigma \mu_{1} } D^{ \mu _{2} }
\cdots D^{ \mu _{n-2} }
\}^{a}
\partial^{\mu_{n-1}}\overline{\chi}^{a} 
- {\rm traces}$.

Again, the operators (\ref{eqn:1})-(\ref{eqn:4}) are not all
independent:
Firstly, due to the Bose statistics of the gluons,
we obtain the symmetry relations
$T_{n,l}^{ \sigma \mu_{1} \cdots \mu_{n-1} }
= T_{n,n-l}^{ \sigma \mu_{1} \cdots \mu_{n-1} }$ and 
$\widetilde{T}_{n,l}^{ \sigma \mu_{1} \cdots \mu_{n-1} }
=-\widetilde{T}_{n,n-l}^{ \sigma \mu_{1} \cdots \mu_{n-1} }$.
Therefore, we can choose $T_{n,l}$ and $\widetilde{T}_{n,l}$
for $l= 2, \cdots, \frac{n-1}{2}$ as independent operators,
as indicated in (\ref{eqn:2}) and (\ref{eqn:2t}).
Secondly, one can derive the relation
between $T_{n,l}$ and $\widetilde{T}_{n,l}$:
\begin{eqnarray}
\widetilde{T}_{n,j}^{ \sigma \mu_{1} \cdots \mu_{n-1} }
&=& \sum_{l=2}^{j-1} \left( C_{l-2}^{n-2-j} - C_{l-2}^{j-2} \right)
(-1)^{l}T_{n,l}^{ \sigma \mu_{1} \cdots \mu_{n-1} } 
+ \left( C_{j-2}^{n-2-j} -2 \right) (-1)^{j}
T_{n,j}^{ \sigma \mu_{1} \cdots \mu_{n-1} }
\nonumber \\
&+& \sum_{l=j+1}^{(n-1)/2} \left( C_{l-2}^{n-2-j} - C_{n-2-l}^{n-2-j} 
\right)
(-1)^{l}T_{n,l}^{ \sigma \mu_{1} \cdots \mu_{n-1} },
\label{eq:rel2}
\end{eqnarray} 
where $C_{r}^{n}=n!/[ r!(n - r)! ]$. This relation is also 
due to the Bose statistics of the gluons.
Thirdly, the analogue of (\ref{operel}),
relating the gluon bilinear operator with the trilinear ones,
can be obtained as
\bea
T_{n,G}^{ \sigma \mu_{1} \cdots \mu_{n-1} }
=
\sum^{\frac{n-1}{2}}_{l = 2 }
\left[ 
\frac{( l - 2 )}{2(n-1)}C^{n -2}_{l}-
\frac{( n - l - 2 )}{2(n-1)}C^{n-2}_{n-l}
-\frac{n(n-2l)}{2(n-2)(n-1)}C^{n-2}_{l-1}
+ (-1)^{l+1} 
\right](-1)^{l}
\nonumber \\
\hspace*{-1pc}
\times 
T_{n,l}^{ \sigma \mu_{1} \cdots \mu_{n-1} }
+
\frac{1}{n-1}T_{n,E}^{ \sigma \mu_{1} \cdots \mu_{n-1}}
+
\frac{1}{n-1}T_{n,B}^{ \sigma \mu_{1} \cdots \mu_{n-1}}
+ 
\frac{n}{n-1} \sum^{n-2}_{l = 1} C^{n - 3}_{l-1}
(-1)^{l+1} 
R^{ \sigma \mu_{1} \cdots \mu_{n-1}}_{n,l}.
\label{eq:rel3}
\eea
Note that the 
quark-gluon-quark operator
$R_{n,l}$ of (\ref{eq:rl})
as well as the operators
(\ref{eqn:2}), (\ref{eqn:3}) and (\ref{eqn:4})
appears in the r.h.s.
To derive this relation, 
we have used $[D_{\mu}, D_{\nu}] = -ig G_{\mu \nu}$
and the identities:
\begin{equation}
D_{\sigma}G_{\nu \alpha} + D_{\nu} G_{\alpha \sigma} 
+ D_{\alpha} G_{\sigma \nu} = 0; \;\;\;\;
D_{\sigma}\widetilde{G}_{\nu \alpha} + D_{\nu} 
\widetilde{G}_{\alpha \sigma} 
+ D_{\alpha} \widetilde{G}_{\sigma \nu} = 
\varepsilon_{\nu \alpha \sigma \rho} D_{\lambda} G^{\lambda \rho},
\label{eq:jli}
\end{equation}
where the first identity is the usual Bianchi identity
while the second one is a consequence of the relation
$g_{\mu \nu} \varepsilon_{\alpha \beta \gamma \delta}
= g_{\mu \alpha} \varepsilon_{\nu \beta \gamma \delta}
+ g_{\mu \beta} \varepsilon_{\alpha \nu \gamma \delta}
+ g_{\mu \gamma} \varepsilon_{\alpha \beta \nu \delta}
+ g_{\mu \delta} \varepsilon_{\alpha \beta \gamma \nu}$.
The operator identity (\ref{eq:rel3}) is new,
and is one of the main results of this work.
As a result of (\ref{eq:rel2}), (\ref{eq:rel3}), 
we can conveniently choose a set of independent operators as
(\ref{eqn:2}), (\ref{eqn:3}) and (\ref{eqn:4}).

To summarize our independent basis of the twist-3 flavor singlet
operators for the $n$-th moment
(see (\ref{eq:rl})-(\ref{eq211}),  
(\ref{eqn:2}), (\ref{eqn:3}) and (\ref{eqn:4})):
The quark-gluon-quark operators $R_{n,l}$ ($l = 1, \cdots, n-2$);
the quark EOM operator $R_{n,E}$; the quark mass operator
$R_{n,m}$; the three-gluon operators $T_{n,l}$ 
($l = 2, \cdots, \frac{n-1}{2}$); the gluon EOM operator
$T_{n,E}$; the BRST-exact operator $T_{n,B}$.
Among them, $R_{n,l}, R_{n,E}, R_{n,m}$, and $T_{n,l}$ are
gauge-invariant, while $T_{n,E}$ and $T_{n,B}$
are not gauge-invariant but BRST-invariant.
These $(3n+1)/2$ operators 
will mix with each other under renormalization.

\section{Renormalization of twist-3 singlet operators}

\vspace{1mm}
\noindent
The $Q^{2}$-evolution of the flavor singlet part
of $\widetilde{g}_{2}(x, Q^{2})$ is governed
by the anomalous dimensions,
which enter into the RG equation for
the relevant twist-3 singlet operators obtained in sect.2.
In this section we discuss 
the renormalization of these operators to obtain the 
anomalous dimensions.

We follow the standard method to renormalize
the local composite operators\cite{JCOL1}.
We multiply the operators discussed in sect.2 
by a light-like vector
$\Delta_{\mu_{i}}$ to symmetrize the Lorentz indices and to
eliminate the trace terms:
%and denote the results by
$\Delta_{\mu_{1}}\cdots
\Delta_{\mu_{n-1}} R_{n,l}^{\sigma\mu_{1}\cdots \mu_{n-1}}
\equiv
\Delta \cdot R^{\sigma}_{n,l}$,
$\Delta_{\mu_{1}}\cdots
\Delta_{\mu_{n-1}} T_{n,l}^{\sigma\mu_{1}\cdots \mu_{n-1}}
\equiv
\Delta \cdot T^{\sigma}_{n,l}$, etc.
We then embed the operators
${\cal O}_{j}
= \Delta \cdot R^{\sigma}_{n,l},
\Delta \cdot T^{\sigma}_{n,l}$, etc.
into the three-point function as
$\langle 0|{\rm T} {\cal O}_{j}(0) A_{\mu}(x)\psi(y)
\overline{\psi}(z)|0 \rangle$,
$\langle 0|{\rm T} {\cal O}_{j}(0) A_{\mu}(x)A_{\nu}(y)
A_{\rho}(z)|0 \rangle$, etc., and compute the
1-loop corrections.
We employ the Feynman gauge ($\alpha = 1$) and
renormalize the operators in the MS scheme.
To perform the renormalization in a consistent manner without subtle
infrared singularities\cite{JCOL2}, we keep the quark and gluon external
lines off-shell; in this case the EOM operators as well as the BRST-exact 
operators mix through renormalization as nonzero operators.

One serious problem in the calculation is the mixing 
of the many gauge-noninvariant 
as well as BRST-noninvariant 
EOM operators. 
As explained
in \cite{KOD1,KA}, gauge-noninvariant EOM
operators are given by replacing some of
the covariant derivatives $D^{\mu_i}$
by the ordinary derivatives $\partial^{\mu_i}$ in (\ref{eq211}).
Furthermore, in the present case, the BRST-noninvariant
EOM operators obtained similarly from the gluon EOM
operator (\ref{eqn:4}) will also participate in the mixing.
However, the problem can be overcome by direct generalization of the
method employed in \cite{KOD2,TK}.
We introduce the vector $\Omega^{r}_{ \mu }$
($r= 1, 2, 3$) satisfying  $\Delta^{ \mu }
\Omega^{r}_{ \mu }=0$ for each external gluon line, and contract the
Green's functions as 
$\Omega^{1}_{\mu}
\langle 0|{\rm T} {\cal O}_{j}(0) A^{\mu}(x)\psi(y)
\overline{\psi}(z)|0 \rangle$,
$\Omega^{1}_{ \mu }\Omega^{2}_{ \nu }\Omega^{3}_{ \rho }
\langle 0 | {\rm T} {\cal O}_{j}(0) 
	A^{ \mu }(x) A^{ \nu }(y) A^{ \rho }(z) | 0 \rangle$, etc. 
%(In principle, one can introduce 
%the two independent $\Omega_{ \mu }$'s
%for this purpose.)
%By this projection, the tree vertex of the gauge non-invariant EOM
%operators coincides with the corresponding gauge invariant one, and
%thus the former never produces new contribution.
This brings the two merits: Firstly,
the tree vertices of the gauge (BRST) invariant 
and the gauge (BRST) noninvariant 
EOM operators coincide. Thus, essentially only one quark (gluon) EOM
operator is now involved in the operator mixing. 
Secondly, the structure of the vertices for the twist-3 operators
are simplified extremely, and the computation becomes more tractable.
%Thirdly, the three-gluon vertex of the BRST-exact operator (\ref{eqn:3})
%vanishes when appropriate additional conditions are imposed among
%$\Omega^{r}_{\mu}$'s. Thus, we can exclude the BRST-exact operator
%from the operator mixing
%in the gluon three-point functions, and treat its mixing separately. 

\begin{figure}[H]
\begin{center}
%\begin{tabular}{cccccc}
\begin{tabular}{ccccccccccc}
\leavevmode\psfig{file=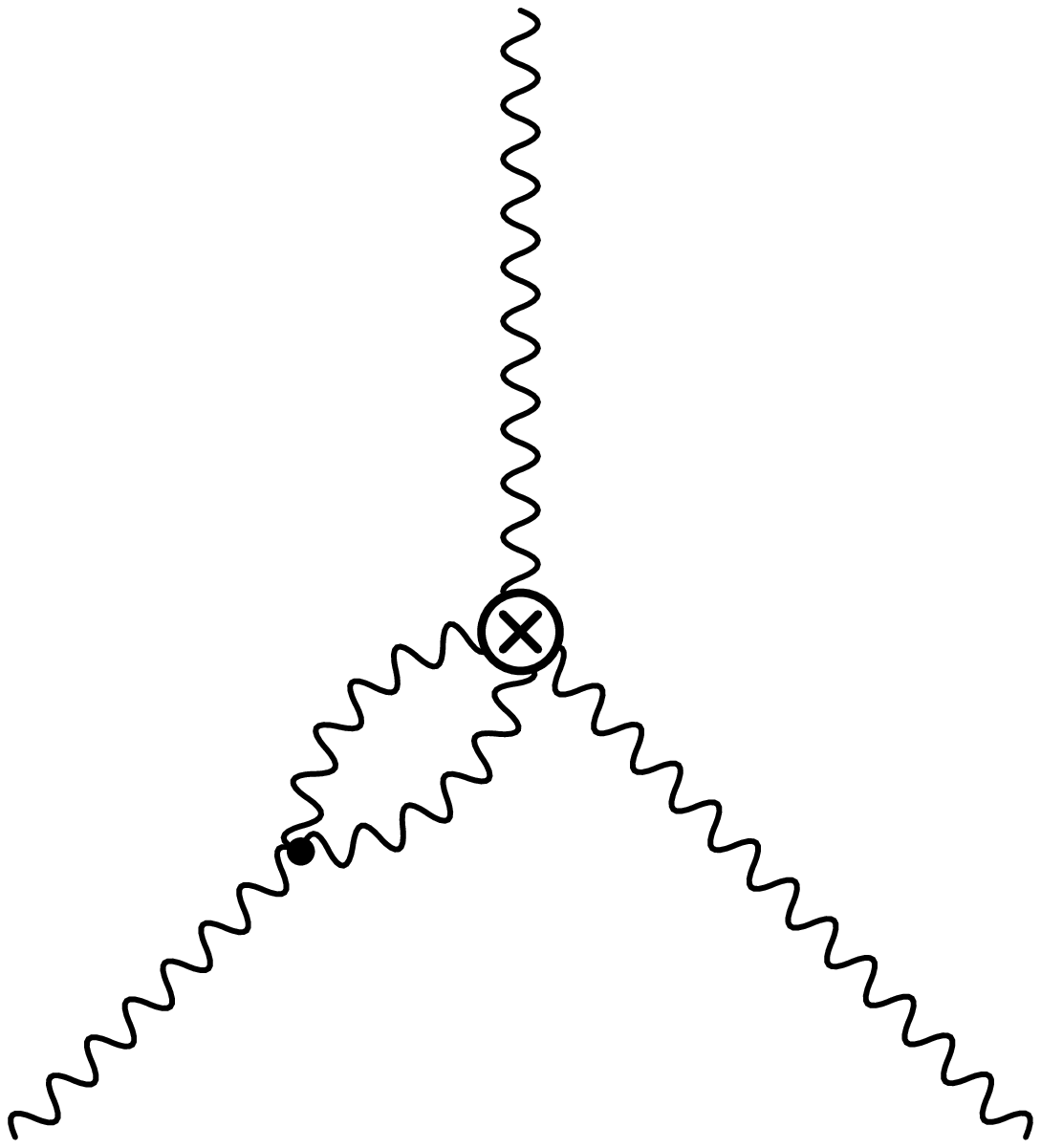,width=1.3cm} &
%%%%%%%%%%%%%%%%%%%%%%%%%%%%
\leavevmode\psfig{file=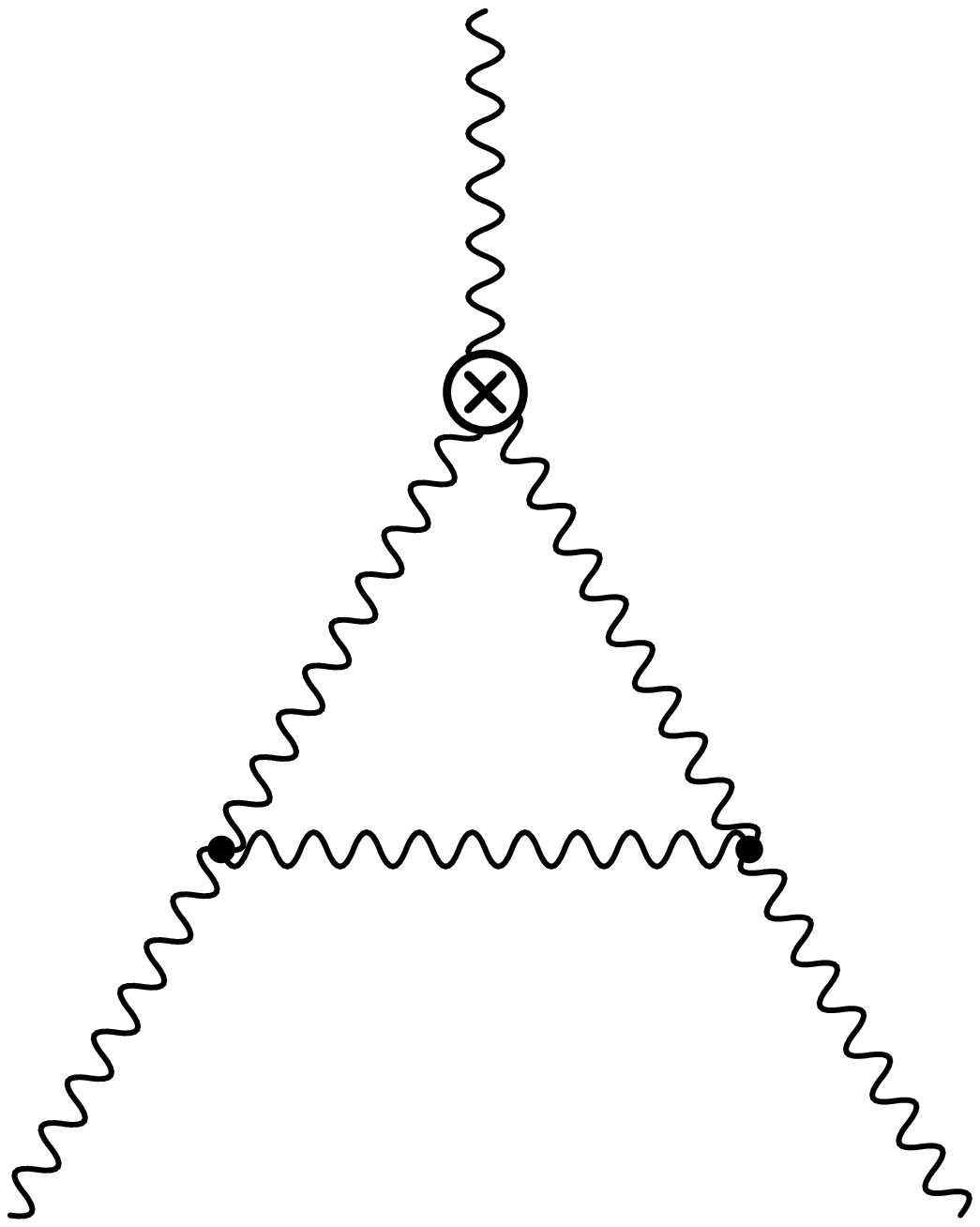,width=1.3cm} &
%%%%%%%%%%%%%%%%%%%%%%%%%%%%
\leavevmode\psfig{file=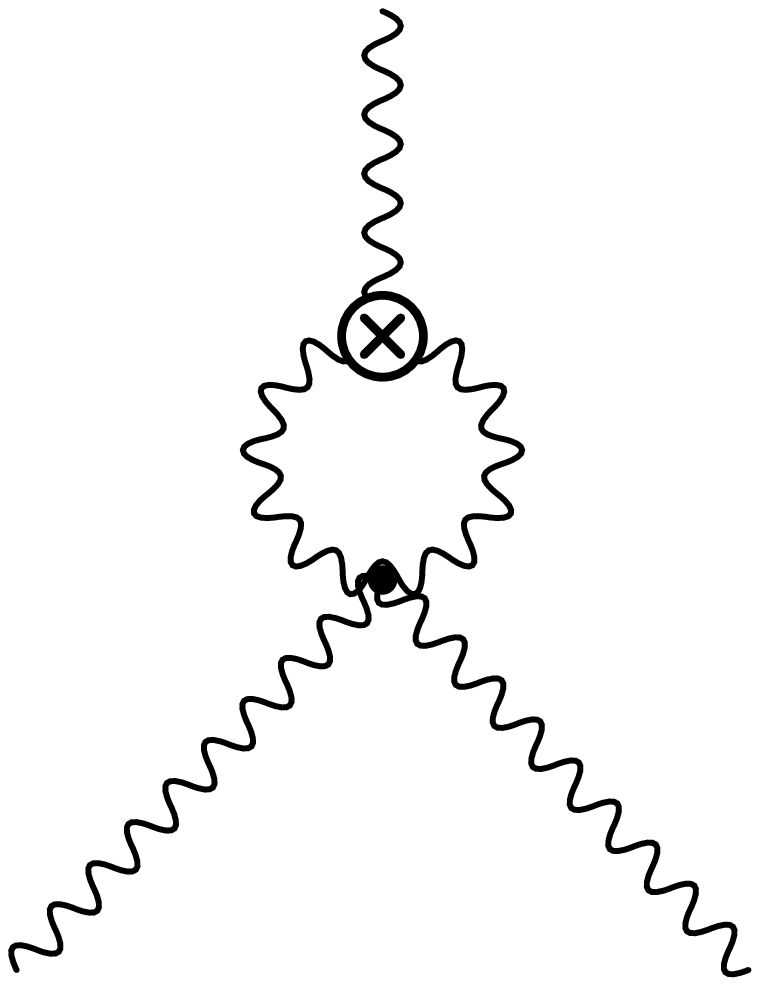,width=1.1cm} & 
%%%%%%%%%%%%%%%%%%%%%%%%%%%%
\leavevmode\psfig{file=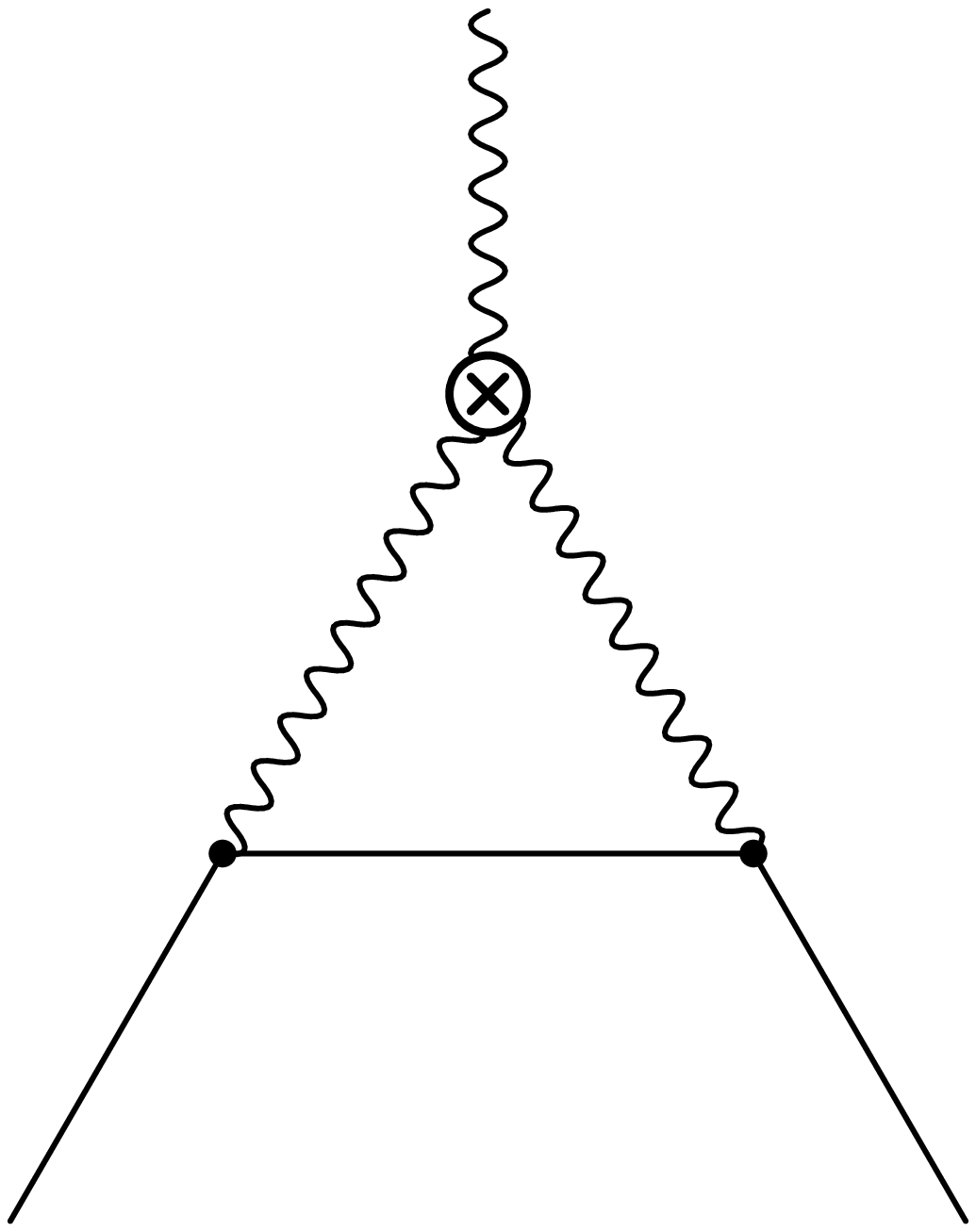,width=1.3cm} &
%%%%%%%%%%%%%%%%%%%%%%%%%%%%
\leavevmode\psfig{file=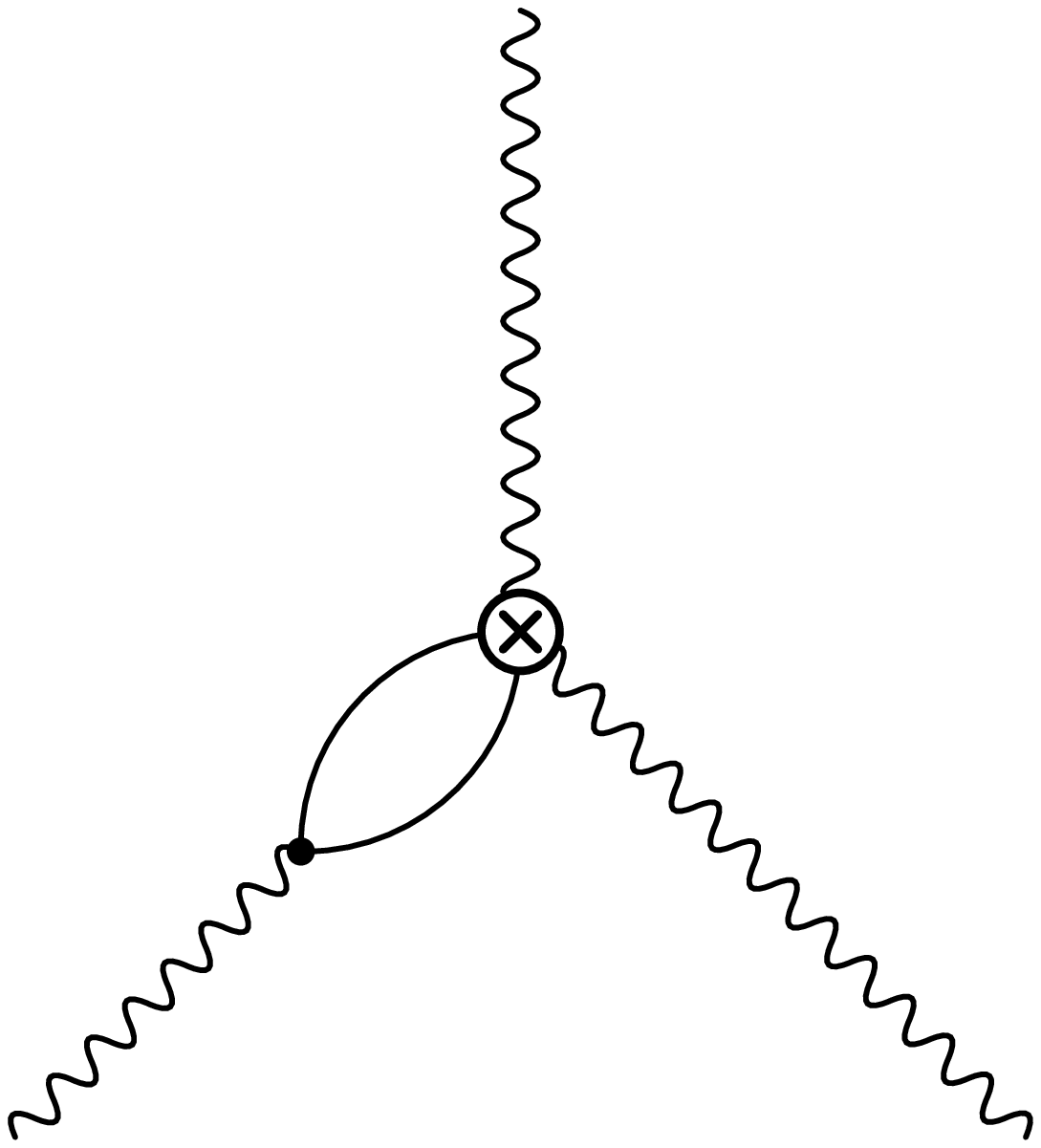,width=1.3cm} &
%%%%%%%%%%%%%%%%%%%%%%%%%%%%
\leavevmode\psfig{file=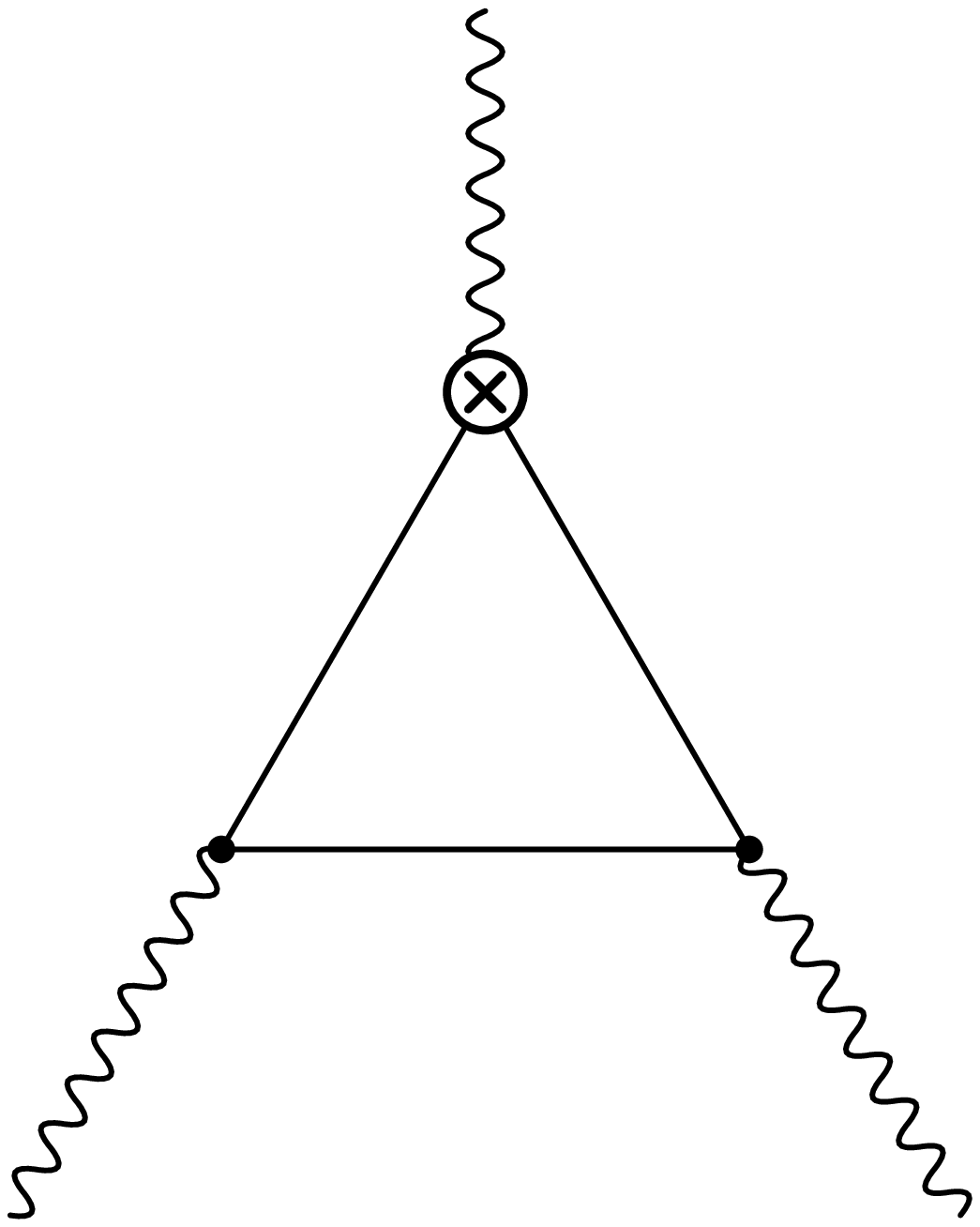,width=1.3cm} &
%\\
%%%%%%%%%%%%%%%%%%%%%%%%%%%%
%(a)&(b)&(c)&(d)&(e)&(f) \\
%%%%%%%%%%%%%%%%%%%%%%%%%%%%
\leavevmode\psfig{file=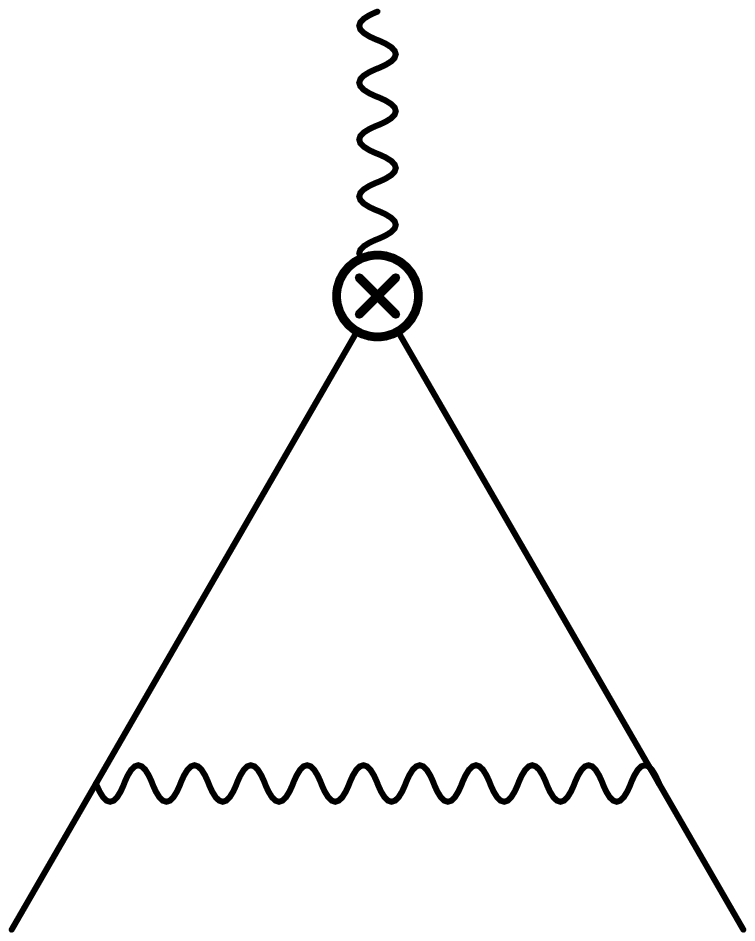,width=1.1cm} &
%%%%%%%%%%%%%%%%%%%%%%%%%%%%
\leavevmode\psfig{file=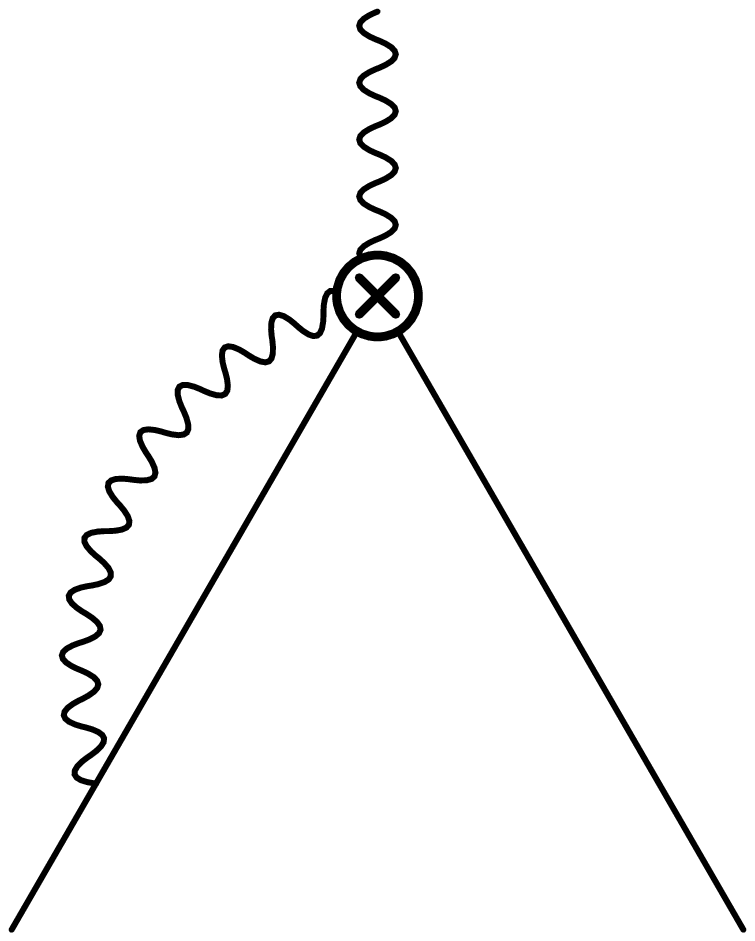,width=1.1cm} &
%%%%%%%%%%%%%%%%%%%%%%%%%%%%
\leavevmode\psfig{file=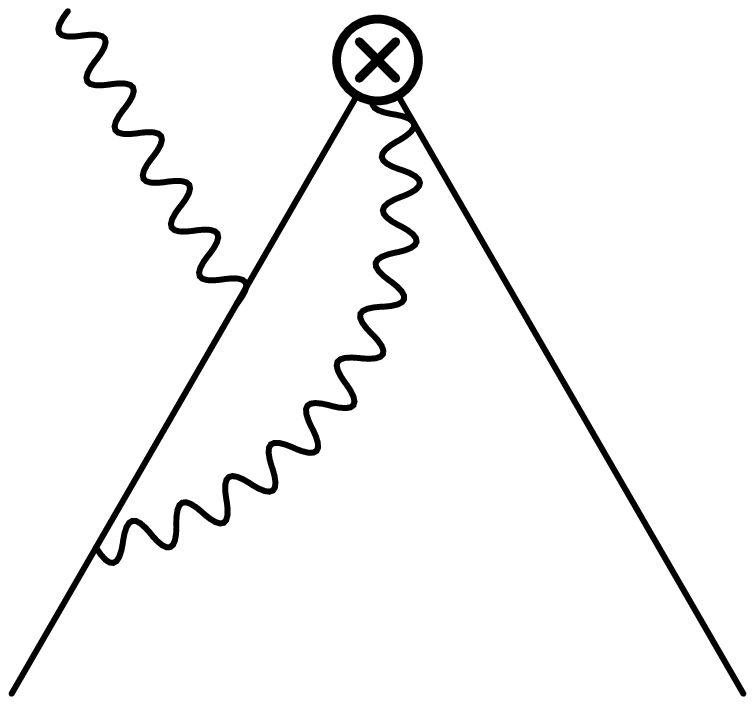,width=1.1cm} &
%%%%%%%%%%%%%%%%%%%%%%%%%%%%
\leavevmode\psfig{file=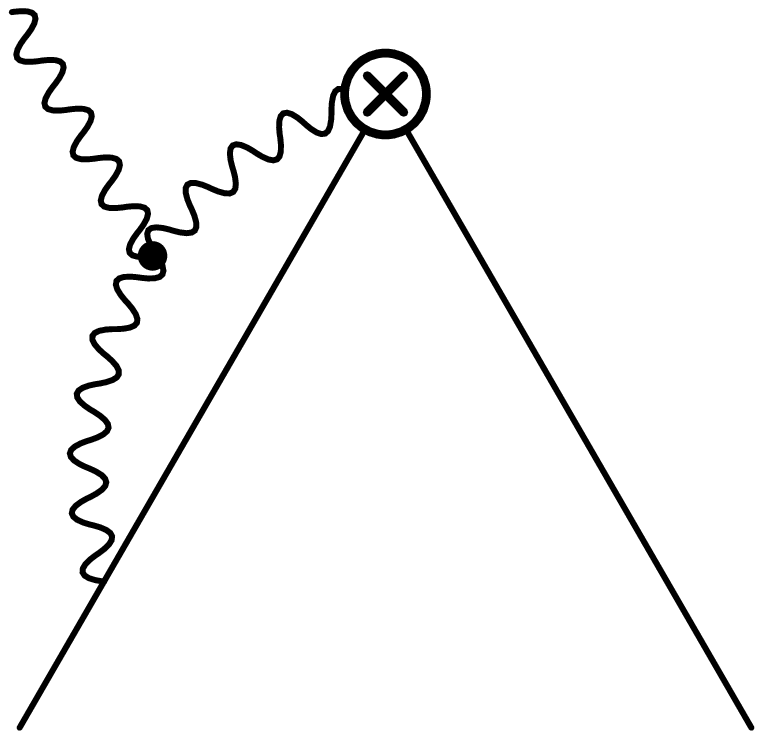,width=1.1cm} &
%%%%%%%%%%%%%%%%%%%%%%%%%%%%
\leavevmode\psfig{file=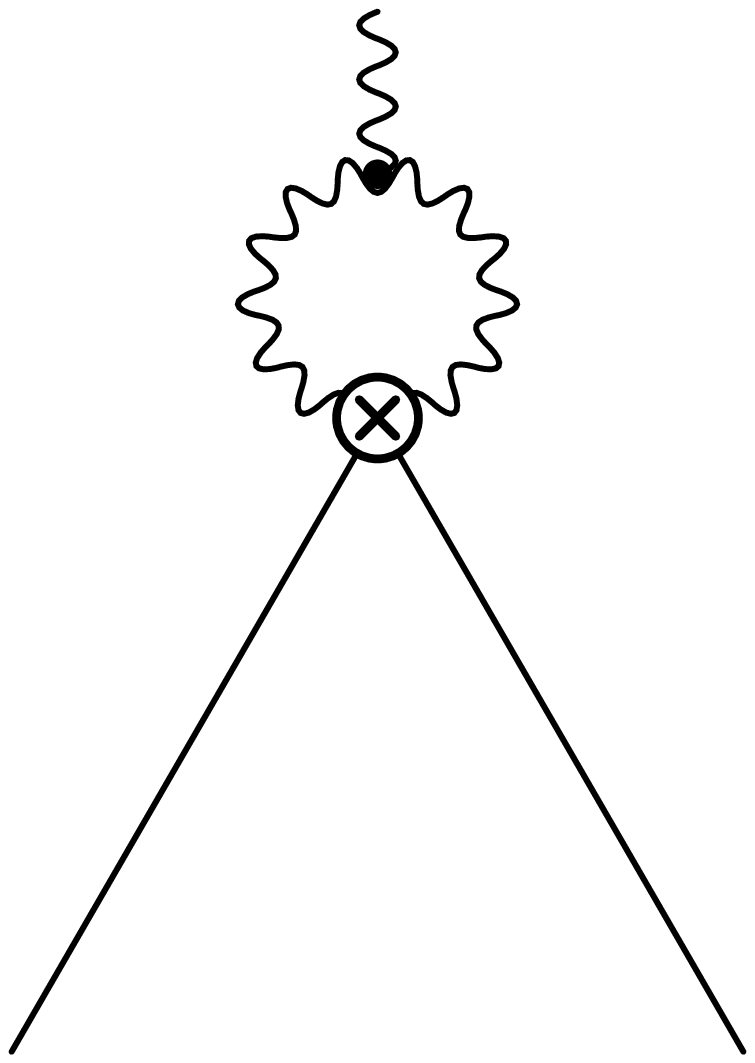,width=1.1cm} \\
%%%%%%%%%%%%%%%%%%%%%%%%%%%%
(a)&(b)&(c)&(d)&(e)&(f) &
(g)&(h)&(i)&(j)&(k)
\end{tabular}
\caption{One-loop diagrams (full lines: quarks; wavy lines: gluons)}
\label{fig:DGM}
\end{center}
\end{figure}

We perform the 1-loop calculation for the one-particle-irreducible
three-point function with the insertion of 
$\Delta \cdot R^{ \sigma }_{n,l}$,
$\Delta \cdot T^{ \sigma }_{n,l}$, etc.
Fig.\ref{fig:DGM} shows the relevant 1-loop diagrams.
\vspace*{-1.5pc}
\hspace{-2mm}
\vspace*{-2pc}
\hspace*{-2pc}
The results can be summarized in the following matrix form:
\bea
\left(
\begin{array}{l}
\Delta\cdot R_{n,i}^{\sigma}\\ 
\Delta\cdot R_{n,m}^{\sigma} \\ 
\Delta\cdot T_{n,l}^{\sigma} \\ 
\Delta\cdot T_{n,B}^{\sigma} \\ 
\Delta\cdot R_{n,E}^{\sigma} \\ 
\Delta\cdot T_{n,E}^{\sigma}
\end{array}
\right)_{bare}
~=~
\left(
\begin{array}{cccccc}
Z_{ik}^{FF} & Z_{im}^{FF} & Z_{ij}^{FG} & Z_{iB}^{FG} & Z_{iE}^{FF} &
Z_{iE}^{FG} \\
0 & Z_{mm}^{FF} & 0 & 0 & 0 & 0 \\
Z_{lk}^{GF} & Z_{lm}^{GF} & Z_{lj}^{GG} & Z_{lB}^{GG} & Z_{lE}^{GF} &
Z_{lE}^{GG} \\
0 & 0 & 0 & Z_{BB}^{GG} & Z_{BE}^{GF} & Z_{BE}^{GG} \\
0 & 0 & 0 & 0 & Z_{EE}^{FF} & Z_{EE}^{FG} \\
0 & 0 & 0 & 0 & Z_{EE}^{GF} & Z_{EE}^{GG} 
\end{array}
\right)
\left(
\begin{array}{l}
\Delta\cdot R_{n,k}^{\sigma} \\ 
\Delta\cdot R_{n,m}^{\sigma} \\ 
\Delta\cdot T_{n,j}^{\sigma} \\ 
\Delta\cdot T_{n,B}^{\sigma} \\ 
\Delta\cdot R_{n,E}^{\sigma} \\ 
\Delta\cdot T_{n,E}^{\sigma}
\end{array}
\right),
\label{eqn:6}
\eea
where the operators with (without) the suffix ``$bare$'' are the bare
(renormalized) quantities.
In the MS scheme we express the renormalization constants
$Z_{pp'}^{AA'}$ as
\bea
Z_{pp'}^{AA'}~\equiv~ \delta_{pp'} \delta_{AA'} + 
\frac{g^{2}}{8 \pi^{2}(4-D)} X_{pp'}^{AA'}
\label{eqn:7}
\eea
with $D$ the space-time dimension.
Here $p=1,\cdots,n-2,m,E$ for $A=F$, while
$p=2,\cdots,(n-1)/2,B,E$ for $A=G$.
The renormalization constant matrix is triangular because the physical 
matrix elements of the EOM operators and of the BRST-exact operators
vanish\cite{JCOL1}.

We here concentrate on diagrams (e)-(k) which are relevant for an
example discussed in sect.4:
Diagrams (g)-(k) have already been computed in the present scheme, and 
the results have been obtained as $Z_{ik}~(i,k=1,\cdots,n-2)$ of
eqs.(14)-(17) in \cite{KOD2}.
In the singlet case, however, these renormalization constants $Z_{ik}$,
which we denote by $(Z_{ik})_{NS}$ here and in the following, 
should be identified as 
%(see eqs.(\ref{eqn:4}),(\ref{eqn:6}))
%
\bea
(Z_{ik})_{NS}~=~Z_{ik}^{FF} + (-1)^{k} n C_{k-1}^{n-3} Z_{iE}^{FG},
\label{eqn:8}
\eea
because $T_{n,E}$ of (\ref{eqn:4}) contains the quark-gluon-quark
term $i^{n-1} {\cal S}\{ \widetilde{G}^{ \sigma \mu_{1} } D^{ \mu _{2} }
\cdots D^{ \mu _{n-2} }
\}^{a}
g \overline{ \psi } t^{a} \gamma^{ \mu_{n-1} } \psi$,
which can be reexpressed by $R_{n,k}$ with the coefficient
in the second term of (\ref{eqn:8}).
On the other hand, 
$Z_{im},Z_{iE}$ and $Z_{mm}$ given by eq.$(18)$ of \cite{KOD2}
should be identified as $Z_{im}^{FF},Z_{iE}^{FF}$ and $Z_{mm}^{FF}$.
From the computation of diagrams (e) and (f), we obtain 
%$Z_{lj}^{FG},~Z_{lE}^{FG}$ as
\begin{equation}
X_{lE}^{FG} 
	= (-1)^{l}
\frac{N_{f}}{n^{2}}
\frac{4}{C^{n - 1}_{l}},
\label{eqn:12}
\end{equation}
where $N_{f}$ is the number of quark flavors,
and $l=1,\cdots,(n-1)/2$.
For $l=(n-1)/2+1,\cdots,n-2$, the replacement $l \rightarrow l'=n-l-1$ 
should be understood.
(\ref{eqn:8}) and (\ref{eqn:12}) determine $Z_{ik}^{FF}$
for general $n$, and the result agrees with that of \cite{BKL}.
%\cite{BKL,DMU,BB}.
Similar computation including other 
diagrams (a)-(d) gives other renormalization constants.
These results will be
published elsewhere.

The RG equation for the twist-3 singlet operators
can be obtained in a standard way from (\ref{eqn:6}).
As is well known, the RG equation decouples for each $n$.
In the leading-logarithmic approximation (LLA),
this equation is solved to give
\begin{eqnarray}
\langle \langle R_{n,i}(Q^{2}) \rangle \rangle
&=& \sum_{k=1}^{n-2} \langle\langle R_{n,k}(\mu^{2}) \rangle \rangle
\left[ L^{\frac{X}{b}} \right]^{FF}_{ik} 
+\langle\langle R_{n,m}(\mu^{2}) \rangle \rangle
\left[ L^{\frac{X}{b}} \right]^{FF}_{im}
+\sum_{j=2}^{\frac{n-1}{2}}
\langle\langle T_{n,j}(\mu^{2}) \rangle \rangle
\left[ L^{\frac{X}{b}} \right]^{FG}_{ij} \label{eq:sol1} \\
\langle \langle R_{n,m}(Q^{2}) \rangle \rangle
&=& \langle\langle R_{n,m}(\mu^{2}) \rangle \rangle
L^{X_{mm}^{FF}/b}, \label{eq:sol2} \\
\langle \langle T_{n,l}(Q^{2}) \rangle \rangle
&=& \sum_{k=1}^{n-2} \langle\langle R_{n,k}(\mu^{2}) \rangle \rangle
\left[ L^{\frac{X}{b}} \right]^{GF}_{lk} 
+\langle\langle R_{n,m}(\mu^{2}) \rangle \rangle
\left[ L^{\frac{X}{b}} \right]^{GF}_{lm}
+\sum_{j=2}^{\frac{n-1}{2}}
\langle\langle T_{n,j}(\mu^{2}) \rangle \rangle
\left[ L^{\frac{X}{b}} \right]^{GG}_{lj} \label{eq:sol3}
\end{eqnarray}
with $i = 1, \cdots, n-2$, $l=2, \cdots, \frac{n-1}{2}$,
$L = \alpha_{s}(\mu^{2})/\alpha_{s}(Q^{2})$, and 
$b=(11N_{c}-2 N_{f})/3$ for $N_{c}$ color.
Here we introduced the reduced matrix element 
$\langle \langle {\cal O}(\mu^{2}) \rangle \rangle$
by
\begin{equation}
\langle P S | {\cal O}^{\sigma \mu_{1} \cdots \mu_{n-1}}
(\mu^{2}) |PS\rangle = -2
\langle \langle {\cal O}(\mu^{2}) \rangle \rangle
{\cal S}{\cal A} \left[ S^{\sigma} P^{\mu_{1}} \cdots
P^{\mu_{n-1}} \right] - {\rm traces},
\label{eq:rmet}
\end{equation}
with ${\cal O} = R_{n,i}, R_{n,m}$, and $T_{n,l}$.
(\ref{eq:sol1})-(\ref{eq:sol3})
exhibit the complicated mixing
characteristic of the higher-twist operators.
On the other hand, the EOM operators $R_{n,E}$, $T_{n,E}$
as well as the BRST-exact operator $T_{n,B}$ decouple
from the result
%(\ref{eq:sol1})-(\ref{eq:sol3})
because their matrix elements vanish\cite{JCOL1}.
 
Before ending this section, we give explicit formula
for the $Q^{2}$-evolution of $g_{2}(x, Q^{2})$ in the LLA:
In the LLA, we substitute, in (\ref{eq:g2}),
$\mu^{2} = Q^{2}$ and
$H_{2i}\left(\frac{x}{\xi}, 1, \alpha_{s}(Q^{2})\right)
= (e^{2})_{ii}
\delta \left( \frac{x}{\xi} -1 \right) 
+ O(\alpha_{s})$ for $i = u, \overline{u}, d, \overline{d}, \cdots$,
with $e$ the charge matrix of the quarks. 
Therefore, the $Q^{2}$-evolution
of $g_{2}(x, Q^{2})$ is governed directly
by the scale dependence of the local composite operators
corresponding to the moment of $\phi^{i}_{2}(\xi, Q^{2})$.
As shown in (\ref{eq:ww}),
the twist-2 contributions contained in $g_{2}(x, Q^{2})$
are determined completely by $g_{1}(x, Q^{2})$
whose $Q^{2}$-evolution is well known.
On the other hand, for the twist-3 part 
$\widetilde{g}_{2}(x, Q^{2})$,
we have already determined an independent basis 
of relevant operators
and their scale dependence.
%obtained the relevant local operators in sect.2.
Corresponding to the decomposition of the charge matrix 
squared by $e^{2} = \langle e^{2} \rangle \mbox{\boldmath $1$}
+ A_{3} \lambda_{3} + A_{8} \lambda_{8}$ with $\lambda_{i}$ 
the flavor matrices,
we decompose $\widetilde{g}_{2}(x, Q^{2})$ into the singlet
and the nonsinglet parts by
$\widetilde{g}_{2}(x, Q^{2}) =
\widetilde{g}_{2}^{S}(x, Q^{2}) + \widetilde{g}_{2}^{NS}(x, Q^{2})$.
Then, from (\ref{eq:g2})-(\ref{operel}) and (\ref{eq:rmet}),
it is straightforward to see that 
the moment
$M_{n}[\widetilde{g}_{2}^{S}(Q^{2})]
= \int_{0}^{1} dx x^{n-1} \widetilde{g}_{2}^{S}(x, Q^{2})$
is given by the matrix elements of our independent operators as 
(see also \cite{KOD1})
\begin{equation}
M_{n}[\widetilde{g}_{2}^{S}(Q^{2})]
= \frac{\langle e^{2} \rangle}{2}\left(
\sum_{k=1}^{n-2} (n-1-k) \langle \langle R_{n,k}(Q^{2}) 
\rangle \rangle
+\frac{n-1}{n} \langle \langle R_{n,m}(Q^{2})\rangle \rangle 
\right).
\label{eq:mom}
\end{equation}
The equations (\ref{eq:sol1})-(\ref{eq:mom}) completely
determine the $Q^{2}$-evolution of the singlet part.

The moment of the flavor nonsinglet part
$M_{n}[\widetilde{g}_{2}^{NS}(Q^{2})]$
is given by the equation similar to (\ref{eq:mom})
with $R_{n,k}(Q^{2}), R_{n,m}(Q^{2})$
and $\langle e^{2} \rangle$ replaced by the corresponding 
nonsinglet quantities.
In this case, the relevant scale dependence
is given by
(\ref{eq:sol1}),(\ref{eq:sol2}) with the replacement
$X^{FF} \rightarrow (X)_{NS}$ and 
$\langle \langle T_{n,j}(\mu^{2}) \rangle \rangle \rightarrow 0$.
For the detail
of the nonsinglet part, 
we refer the readers to e.g. \cite{KOD1,KOD2}.

\section{Application to the $n=3$ case}

\begin{figure}[H]
\begin{center}
\leavevmode\psfig{file=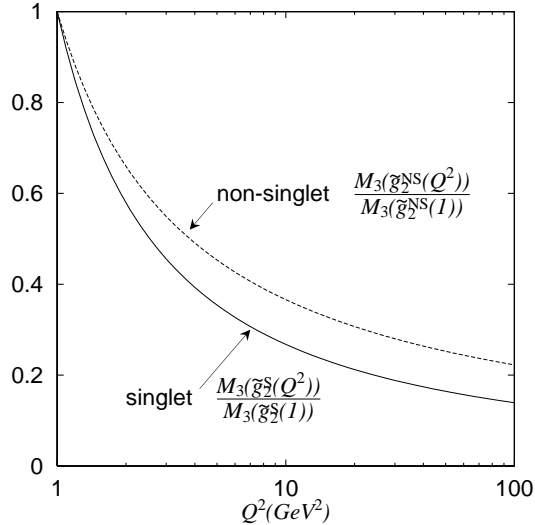,width=8cm,angle=-90}
%%%%%%%%%%%%%%%%%%%%%%%%%%%%
\caption{$Q^{2}$-evolution of $\widetilde{g}_{2}(x, Q^{2})$ for 
the lowest ($n=3$) moment}
\label{fig:q2}
\end{center}
\end{figure}

\vspace{1mm}
\noindent
We now present an example of the $Q^{2}$-evolution for the lowest $(n=3)$
moment.
In this case there exists no three-gluon operator $\Delta\cdot
T_{n,l}^{\sigma}$, while there exists one quark-gluon-quark operator
$\Delta \cdot R_{3,1}^{\sigma} = \frac{1}{3}g \overline{\psi}
\widetilde{G}^{\sigma
\mu} \Delta_{\mu} \slashl{\Delta} \psi$ 
(see (\ref{eqn:2}), (\ref{eq:rl})).
If we neglect the contribution of the quark mass operator 
$\Delta \cdot R_{3,m}^{\sigma}$, 
%which is proportional to the quark mass, 
only one
operator $\Delta \cdot R_{3,1}^{ \sigma }$ contributes to the
$Q^{2}$-evolution of the physical matrix elements
as given by (\ref{eq:sol1}).
Therefore, the $Q^{2}$-evolution is completely determined from our
results (\ref{eqn:8}), (\ref{eqn:12}):
From (\ref{eq:sol1}), the relevant renormalization constant is
$Z_{11}^{FF}$.
We obtain, from (\ref{eqn:8}),(\ref{eqn:12}) and
eqs.(14)-(17) of \cite{KOD2},
$Z_{11}^{FF} = (Z_{11})_{NS} + 3Z_{1E}^{FG} 
	     = 1+
g^{2} \left( C_{F}/3 - 3C_{G} - 2N_{f}/3 \right)/[8 \pi^{2}(4-D)]$,
where $C_{F}=(N_{c}^{2}-1)/2N_{c},~C_{G}=N_{c}$.
This result combined with (\ref{eqn:7}), (\ref{eq:sol1}) and 
(\ref{eq:mom})
gives the $Q^{2}$-evolution:
\bea
%\int_{0}^{1}dx x^{2}g_{2}^{tw.3}(x, Q^{2})
M_{3}\left[ \widetilde{g}_{2}^{S}(Q^{2}) \right] 
=
\left(
\frac{\alpha_{s}(Q^{2})}{\alpha_{s}(\mu^{2})}
\right)^{(3C_{G} - C_{F}/3 + 2N_{f}/3)/b}
%\int_{0}^{1}dx x^{2}g_{2}^{tw.3}(x, \mu^{2}).
M_{3}\left[ \widetilde{g}_{2}^{S}(\mu^{2}) \right].  
\label{eqn:13}
\eea
%
%and $\alpha_{s}(Q^{2})$ is the running
%coupling constant.
Several comments are in order here:
(i) The result (\ref{eqn:13}) coincides with that of \cite{SV,BKL}, 
though our approach is quite different from those works.
This fact confirms the theoretical prediction (\ref{eqn:13}), and
also demonstrates the efficiency of our method.
(ii) The term $2 N_{f} /3$ would be absent from the exponent of
(\ref{eqn:13}), if we considered the nonsinglet case\cite{KOD1,KOD2}.
Our derivation shows that this difference of $2N_{f}/3$ comes from
$Z_{1E}^{FG}$, which describes the mixing between $\Delta \cdot
R_{3,1}^{\sigma}$ and the gluon EOM operator $\Delta \cdot
T_{3,E}^{\sigma}$.
This reveals a peculiar role played by the EOM operator: It
produces observable effects although its physical matrix elements
vanish.
(iii) For $N_{c}=3$ and $N_{f}=4$, the exponent of (\ref{eqn:13}) is
$101/9b$, while the corresponding exponent for the nonsinglet case
is $77/9b$.
Thus the singlet case obeys rather stronger $Q^{2}$-evolution compared 
to the nonsinglet case.
Fig.\ref{fig:q2} shows the behavior of (\ref{eqn:13})
for the case of $N_{c}=3, N_{f} = 4$, and $\Lambda_{QCD} = 0.5$GeV.
\vspace*{-1.5pc}
\vspace*{-3pc}
%
%\hspace*{-2pc}

\section{Conclusions}

\vspace{1mm}
\noindent
In the present study, we have developed
a manifestly covariant approach
to investigate the flavor singlet part of $g_{2}(x,Q^{2})$.
%We employed the framework with the
%manifest Lorentz covariance being kept. 
%We adopted the
%Feynman gauge and dimensional regularization.
We gave a thorough OPE analysis
of the twist-3 flavor singlet operators
which contribute to $g_{2}(x, Q^{2})$.
We derived the operator identities
which relate the two-particle operators
with the three-particle ones.
We have chosen the three-particle 
operators 
%which include the gluon field
%strength explicitly 
as an independent operator's basis.
%although this choice of basis is never compulsory.
To identify the renormalization constants correctly,
the off-shell Green's functions are considered.
We have shown that the EOM operators as well as the 
BRST-exact operators
play an important role.
%to complete
%the renormalization of composite operators.
%and the structure
%of the renormalization constant matrix takes the triangular
%form expected from the general argument \cite{COLL}. 
As an application, we computed the $Q^{2}$-evolution
of the lowest ($n=3$) moment. Our result confirmed
the prediction obtained by different methods\cite{SV,BKL}.

%If we could calculate the \lq\lq on-shell\rq\rq matrix elements
%of composite operators in terms of purely perturbative Feynman
%graphs, we could obtain the enough informations without
%considering the EOM operators. 
%Our calculation based on 
The off-shell Green's
functions are free from the infrared singularity coming from
the collinear configuration, as well as from any unphysical
singularities.
%can not be regulated \cite{EFP}.
We believe that calculating the off-shell Green's functions
is the safest method to obtain the anomalous dimensions.
%We hope that future precise measurements on $g_{2}(x, Q^{2})$ will
%give opportunities to test our results and to obtain information of
%the quark-gluon correlation in the nucleon. 
%We expect that future presice measurements on $g_2(x, Q^{2})$ will
%clarify the effect of twist-3 operators which may be
%the first quantity to see the higher-twist effect in QCD.

%\section*{Acknowledgment}
%
%\vspace{1mm}
%\noindent
%
\vspace{5mm}
\noindent
{\Large\bf Acknowledgment}

\vspace{1mm}
\noindent
The work of J.K. 
was supported in part by the Monbusho
Grant-in-Aid for Scientific Research No. C-09640364.
The work of K.T. 
was supported in part by the Monbusho
Grant-in-Aid for Scientific Research No. 09740215.

\end{document}